\documentclass[preprint, a4paper, 3p, 11pt]{elsarticle}

\usepackage{graphicx}
\usepackage{amssymb}
\usepackage{amsmath}
\usepackage[section]{placeins}

\setcitestyle{authoryear}
\setcitestyle{round}
\setcitestyle{semicolon}

\usepackage{lineno}
\usepackage{url}
\usepackage[]{algorithm2e}
\usepackage{setspace}
\usepackage{braket}
\usepackage{mathrsfs}
\usepackage{booktabs}
\usepackage{multirow} 

\usepackage[dvipsnames]{xcolor}
\makeatletter
\def\ps@pprintTitle{%
 \let\@oddhead\@empty
 \let\@evenhead\@empty
 \def\@oddfoot{}%
 \let\@evenfoot\@oddfoot}
\makeatother

\DeclareMathOperator*{\argmin}{arg\,min}

\begin{document}

\let\today\relax

\title{Evaluating utility in synthetic banking microdata applications}

\author[1,2]{Hugo E. Caceres\corref{corresponding}}
\ead{hcaceresc@bcp.gov.py}
\cortext[corresponding]{Corresponding author}

\author[1,3,4]{Ben Moews}
\ead{ben.moews@ed.ac.uk}

\affiliation[1]{organization={Business School, University of Edinburgh},
    addressline={29 Buccleuch Pl}, 
    city={Edinburgh},
    citysep={}, 
    postcode={EH8 9JS}, 
    country={UK}}

\affiliation[2]{organization={Central Bank of Paraguay},
    addressline={Federación Rusa y Augusto Roa Bastos}, 
    city={Asuncion},
    postcode={001530}, 
    country={Paraguay}}

\affiliation[3]{organization={Centre for Statistics, University of Edinburgh},
    addressline={Peter Guthrie Tait Rd}, 
    city={Edinburgh},
    citysep={}, 
    postcode={EH9 3FD}, 
    country={UK}}

\affiliation[4]{organization={Centre for Financial Innovations, University of Edinburgh},
    addressline={1 Lauriston Pl}, 
    city={Edinburgh},
    citysep={}, 
    postcode={EH3 9EU}, 
    country={UK}}

\date{}

\begin{abstract}

Financial regulators such as central banks collect vast amounts of data, but access to the resulting fine-grained banking microdata is severely restricted by banking secrecy laws. Recent developments have resulted in mechanisms that generate faithful synthetic data, but current evaluation frameworks lack a focus on the specific challenges of banking institutions and microdata. We develop a framework that considers the utility and privacy requirements of regulators, and apply this to financial usage indices, term deposit yield curves, and credit card transition matrices. Using the Central Bank of Paraguay's data, we provide the first implementation of synthetic banking microdata using a central bank's collected information, with the resulting synthetic datasets for all three domain applications being publicly available and featuring information not yet released in  statistical disclosure. We find that applications less susceptible to post-processing information loss, which are based on frequency tables, are particularly suited for this approach, and that marginal-based inference mechanisms to outperform generative adversarial network models for these applications. Our results demonstrate that synthetic data generation is a promising privacy-enhancing technology for financial regulators seeking to complement their statistical disclosure, while highlighting the crucial role of evaluating such endeavors in terms of utility and privacy requirements.

\end{abstract}

\maketitle

\onehalfspacing

\section{Introduction}

Financial regulators such as central banks have a mandate to promote stability and prevent system-wide episodes that can result in crisis. Recent financial turmoil has further elevated their role in financial systems to avoid episodes like the 2007--2008 financial crisis. Since then, there has been an increase in   the legal powers instilled into these institutions when it comes to regulating and supervising the sector, with attribution to request information directly shared from system participants such as banks, investment firms, and other financial entities \citep{masciandaro2018central}. Regulators use the data available to assess the current state of the financial system, identify potential systemic risk, and take early action to avoid financial catastrophe.  

Modern financial systems have not been immune to the exponential growth of data, covering millions of transactions and customers. Regulators modernise their data and reporting requirements, and the resulting information diversity and complexity is considerable, ranging from consolidated financial and capital adequacy reports to fine-grained market and loan transaction data. For example, it is not uncommon for central banks to implement loan-level data requirements that require more than 100 features per product. The Central Bank of Paraguay (CBP) collects at least 60 data points\footnote{https://www.bcp.gov.py/estructura-de-archivos-central-de-informacion-i1102} for every financial operation originated by a banking institution.   

Financial regulators are often pressured to augment their statistical disclosure with fine-grained data, improving market transparency and discipline while also justify the demanding data collection exercises and required resources. However, a central bank's hope to provide utility by opening access fades rapidly when it considers its fiduciary and legal duty to preserve the privacy of the data collected, and the penalties that can result from noncompliance.

Fortunately, the demand for industry applications of privacy-enhancing technologies has led to impressive growth in the fields of computational privacy and statistical disclosure control. Here, a promising technology is synthetic data generation (SDG), which involves computational methods that can generate simulated datasets with privacy protections for many types of data. For financial regulators, SDG presents an attractive pathway to improve statistical disclosure of banking microdata to interested external parties, for example to enable credit risk research.

The controlled release of synthetic data also enables external validation of potential information products. In turn, this helps regulators to mitigate reputational risks associated with information leakage and build trust through transparent data governance. This pathway is of particular interest to regulators in a more incipient stage of microdata disclosure, such as the Central Bank of Paraguay, since the resources required to implement synthetic information products can be lower than maintaining integrated data portals or processing individual information requests. 
 
However, a financial regulator's road to SDG implementation is not straightforward. Evaluation frameworks found in the literature focus on assessing privacy loss, classification accuracy, and dataset characteristics, which works well for general applications but does not provide an accurate representation of the advantages and limitations.

In this work, we provide the, to our knowledge, first implementation of synthetic data generation using a central bank's microdata. Our financial inclusion, term yield curve, and credit risk domain applications cover approximately 3.7 million unique individuals, 13,000 term deposits, and 700,000 credit cards, respectively. We provide a practical case study of SDG for banking microdata, evaluate the utility with domain applications directly related to a central bank's priorities, and identify specific challenges faced by regulators. We also publicly release the resulting synthetic datasets\footnote{The URL to the dataset will be included here in the final version for publication.} through the Zenodo open repository under the European OpenAIRE programme.

\section{Literature Review}

\subsection{The evolution of financial regulation}

The banking system in the United States illustrates the evolution of financial regulation over time. In 1933, the Glass-Steagall Act was passed due to the systemic failure of the sector, followed by the Great Depression. It was thought to impose unreasonable limitations and repealed in 1999. In 2007, the U.S. banking sector collapsed and panic spread to international markets, which has been linked to that decision \citep[see, for example,][]{rahman2012democracy}. As a result, the Dodd-Frank Wall Street Reform and Consumer Protection Act was passed in 2010, placing more regulation on banking institutions and empowering regulators. In 2018, however, banking regulation was again considered too strict, and the Economic Growth, Regulatory Relief, and Consumer Protection Act was passed, which rolled back many of these imposed limits.

In 1996, Paraguay suffered its most significant banking crisis and a new banking law was passed to empower the Central Bank of Paraguay as a regulator. Modification to the law in 2018 increased the CBP's purview over non-banking financial institutions, while banks aggressively look for more growth and efficiency through mergers and acquisitions \citep{deyoung2009mergers}. 

Banking regulation appears to oscillate between relaxed and tight supervision cycles, and financial regulators often use their ample powers when it is too late to avert a crisis. Despite that, the legal powers that are given to financial regulators have only increased over time. Many financial regulators can, for example, request information directly from participants in the financial system such as banks, investment firms, and other financial entities. 

Related research shows that central banks have experienced a steady increase in transparency and independence, which significantly affects outcomes \citep{dincer2013central}. As more aggregated disclosure appears to not improve banking stability, more of the same data disclosure is unlikely to be the panacea, and new types of disclosure need to effectively address what the market can ingest \citep{semenova2012market}. 

\subsection{Accessibility of banking microdata}

The new standard in regulatory reporting is banking microdata, or data at its most granular level on a per-customer or operation basis.  Such data is comprised of independent reports that vary in granularity and update frequency. Balance sheets, for example, provide account-level information on credit portfolios, while granular loan reports detail the individual exposure of debtors.

Banking regulations typically require banks to disclose aggregated statistics and standardised reports to the public to promote market discipline. However, detailed banking data is confidential and strongly protected by regulations, requiring that no information connectable to an individual person or entity, including the use of external data sources, is to be disclosed \citep{bambauer2011tragedy}. In Paraguay, banking law similarly states that no specific information about financial users can be disclosed. There is currently limited and uneven access to representative and high-quality microdata. Public repositories contain few banking datasets and are seldomly updated. Datasets provided through centralised sources such as Bankscope (now Bank Focus) include financial indicators from many institutions, but are centred on panel and aggregated data to evaluate banking sector performance and dynamics \citep{boussemart2019decomposing, du2018data}.

Some institutions have dedicated research centres that provide access through private data sharing agreements. For example, transaction and account microdata from a large Spanish bank has been used to build a proxy for national consumption for the COVID-19 crisis through queries that are executed externally \citep{carvalho2021tracking}. Similar arrangements include the Federal Reserve of St. Louis, which provides access to banking microdata to researchers after registration and approval process. In these cases, there are hurdles to data access for reproducibility. 

The limitation in scope to a single institution is a drawback in these agreements. Data often exists in data silos, where data providers such as credit card operators or financial institutions face privacy concerns and limited incentives to build effective and interoperable sharing mechanisms \citep{yang2023explainable}. Despite an interest in operational research to explore credit scoring models based on real credit card data, current efforts can only source their data from single institutions \citep{akkoc2012empirical, hon2016models, leow2014intensity, leow2016new, djeundje2019dynamic}.

\subsection{Statistical disclosure and privacy in public institutions}

Statistical disclosure control deals with the evaluation of whether data analysis results be safely published. This area focusses on a balance between utility and privacy, with \citet{rastogi2007boundary} noting that ``perfect privacy can be achieved by publishing nothing at all but has no utility, and perfect utility can be obtained by publishing the original data but offers no privacy.'' For public institutions, the central consideration is disclosure risk in terms of how data could be exploited to reveal information about respondents \citep{templ2017statistical}. 

Examples of types of disclosure risks and privacy threats include membership inference, attribute inference, and reconstruction attacks. In addition, financial regulators' data disclosure is not exclusive to trusted parties, so there is an additional burden in which it must be assumed that the information might be used for exploitation \citep{ohm2009broken}. 

Disclosure events impact the most important asset a public financial institution has, which is trust. Regulators have a legal obligation to maintain data confidentiality, but also a moral and ethical obligation to not harm users who provide their data. Even the most optimal implementation of a privacy-preserving mechanism remains imperfect, requiring the assessment which utility significantly outweighs the risk \citep{bambauer2011tragedy, bellovin2019privacy}. While most regulators publish aggregate statistics through template-based reports, some implement a secure environment where user-defined queries are executed on a database after evaluating privacy risks. This commonly involves a rigorous registration process, as well as adherence to the terms and conditions. 

Most financial regulators that process user-defined queries participate in the International Network for Exchanging Experience on Statistical Handling of Granular Data (INEXDA), a cooperative project of central banks and national statistical institutes with the aim of exchanging experiences on the handling of granular data for research purposes \citep{bender2018inexda}. Some member institutions, such as the Bank of Spain and the German Federal Bank, have developed their own financial data labs. One could extend these trends further and come to expect application-specific banking microdata with high quality and privacy guarantees. 

\subsection{From anonymisation to differential privacy}
\label{sec:anonymisation}

Statistical disclosure control methods, as well as privacy-enhancing technologies, have evolved in parallel to other fields such as machine learning. Historically, most data sanitation methods revolved around subtractive techniques where sensitive variables and identifiers are masked (or suppressed) to maintain privacy. For instance, release-and-forget anonymisation includes a collection of anonymisation techniques that are applied before a public data release \citep{ohm2009broken}. 

These techniques distort the original microdata \citep{templ2017statistical}. Examples of data transformations include removing personally identifiable information from the data (suppression), combining specific features into more general ones (generalisation), shuffling/randomisation, and withholding information such as sampling frames \citep{castro2012recent}. However, anonymisation has proven insufficient to preserve privacy, especially when attackers have access to relevant external information. Notable cases include user behaviour data in massive open online courses, information leakage in the Netflix Prize competition dataset using Amazon review data, and the identification of individuals using AOL's published web search query logs \citep{ohm2009broken}. 

Privacy-preserving techniques have emerged to address these limitations, such as minimum subgroup count, or $k$-anonymity. This method uses transformations to guarantee that at least $k$ indistinguishable copies exist of a set of predefined quasi-identifiers \citep{sweeney2002k}. Here, $k$ defines the required transformations until each quasi-identifier has at least $k$ copies, and some institutions do not disclose aggregated statistics on groups with less than $k$ entries. Another technique is microaggregation, which divides a dataset into partitions of at least $k$ records and replaces numeric features with the centroid of each group \citep{domingo2002practical}. 

Eventually, the sharing, preserving, and using of private information led to an increased interest in statistical disclosure control in the field of computer science, where perspectives based on cryptography emerged. Influential contributions include the work of \citet{dwork2006calibrating} and \citet{dwork2014algorithmic} through differential privacy (DP), an attribute realised if, for given parameters, adversaries are not able to learn about a single point in a database (privacy leakage). 

In terms of statistical disclosure, it means that any analysis ran on the output of a differentially private mechanism will yield the same result regardless of a specific record being present in the database. DP offers a new perspective on how much specific, calibrated noise must be added to make each record indistinguishable. Research and industry applications have adopted DP as the state-of-the-art standard, and several national statistics agencies, for example the U.S. Census Bureau and the Australian Bureau of Statistics, are exploring the use of differential privacy in their public data releases \citep{bailie2019abs, abowd2023confidentiality}.

One limitation is that the design is based on interactive queries on a database instead of public data releases \citep{blanco2022critical}. If a privacy budget is spent, a database can ignore further queries, but a release cannot be `turned off' once disclosed. In addition, DP might be too rigorous when it comes to the utility that public statistical disclosure requires. For example, the disclosure-risk avoidance techniques required to satisfy DP, such as noise adding, might alter the data too much, thus reducing utility \citep{domingo2021limits}. A summary of the challenges of differential privacy for government agencies can be found in \citet{drechsler2023differential}.

\subsection{The promise of synthetic data generation}

\citet{rubin1993statistical} argues that statistical disclosure should not share real microdata and highlights the benefits and feasibility of only sharing synthetic data constructed through multiple imputation techniques, using purpose-built mathematical models or algorithms. The records shared are sampled from aggregates and have no links to real microdata. \citep{gadotti2024anonymization}.	

Today, SDG has broad applications, ranging from private release of confidential datasets to augmenting training sets for data-hungry deep learning architectures in computer vision \citep{jordon2022synthetic}. \citet{assefa2020generating} outlines that synthetic data in finance can solve data use restrictions, lack of historical data, and class imbalances, facilitate the training of privacy-preserving machine learning models, and enable data sharing with the research community.

For regulators, this has several characteristics beneficial for statistical disclosure. Some methods now include formal guarantees such as DP, which could benefit both public data releases and secure research environments, and means that more granular data can be released. Several methods have been developed to generate synthetic data in a privacy-aware fashion. The work on Differentially Private Stochastic Gradient Descent (DPSGD) by \citet{song2013stochastic} enables this for neural networks, and Generative Adversarial Networks (GAN) gave rise to, for example Conditional Tabular GANs (CTGAN) by \citet{xu2019modeling} and Private Aggregation of Teacher Ensembles (PATE) CTGANs by \citet{jordon2018pate} to produce tabular data. 

At the same time, approaches based on data distributions have been developed for the same purpose; for example, Marginal-Based Inference (MBI) models use low-dimensional marginal distributions of the original dataset to produce synthetic data with privacy guarantees. Here, Maximum Spanning Trees (MST) select and measure 2-way and 3-way marginals, add noise using a graphical model, and then generate synthetic data by sampling the noisy distributions \citep{mckenna2021winning}. If such data can be used in an effective and secure manner, it offers a cost-effective answer to ongoing challenges of secure research environment solutions.

\subsection{Evaluation frameworks for synthetic data}

Evaluation frameworks for synthetic data generally focus on utility, fidelity, and privacy. Utility evaluation covers the performance of synthetic data applied to general or specific tasks. It is usually measured by comparing a set of performance metrics between the original and synthetic data, such as accuracy, precision, and error terms. Utility for classification can, for example, be evaluated by comparing the accuracy scores of models trained on either the original dataset or synthetic data \citep{bowen2019comparative, tao2021benchmarking}. 

Fidelity evaluation, on the other hand, focusses on assessing the `truthfulness' of synthetic data. Here, the 1-way and 2-way marginal distributions, total variation distance, and correlation metrics can be used to quantify the similarity between datasets\footnote{\url{https://docs.sdv.dev/sdmetrics}}. Lastly, privacy is evaluated to assess whether privacy attacks on synthetic data could result in learning about the original data. An overview of leakage estimation techniques used to quantify the risk of privacy attacks on specific synthetic datasets or generation methods is presented in \citet{gadotti2024anonymization}.

Some frameworks assess mechanisms through benchmarking utility metrics over a wide range of public testing sets \citep[see, for example,][]{tao2021benchmarking}, but this approach is, to the best of our knowledge, not in current use regarding statistical disclosure from public financial institutions, so it remains unclear if SDG offers an improvement. There is, therefore, a need to provide a practical case of synthetic data generation for banking microdata collected by a financial regulator, evaluate the utility with relevant use cases, identify challenges that regulators face, and provide full-scale synthetic datasets for further research to the operational research community and related fields.

\section{Methodology}

\subsection{Data collection and privacy parameters}
\label{sec:mechanism}

This paper uses banking microdata of 21 financial institutions collected by the Central Bank of Paraguay from 2017--2023, in accordance with the terms specified in a corresponding data sharing agreement. The original microdata includes approximately 3.7 million unique individuals for the financial inclusion application, 13,000 term deposits for the term yield curve application, and 700,000 cards for the credit risk application. Details on the variables and data-driven pre-processing strategies for each use case addressed in this work can be found in Section~\ref{sec:use_cases}.

Existing SDG approaches vary considerably, ranging from proprietary to open-source, with and without strong privacy guarantees, and for structured or unstructured data. Regulators have specific concerns regarding statistical disclosure. They should include strong differential privacy guarantees with a tuneable privacy budget parameter. Here, DP is a characteristic such that a randomised algorithm $\Psi$ 
is ($\varepsilon , \delta$)-differentially private if the condition
\begin{eqnarray}
\label{eq:difpriv}
    P(\Psi(D_1) \in S) \leq e^\varepsilon \cdot P(\Psi(D_2) \in S) + \delta,
\end{eqnarray}
where $\varepsilon$ is the maximum distance between queries on two datasets and $\delta$ is the probability of an information leak, holds for two datasets $D_1$ and $D_2$ differing in a single row for any subset of outputs $S$ \citep{dwork2006calibrating}. In other words, a differentially private mechanism produces synthetic data that is indistinguishable from the original with a probability bounded by $\epsilon$ and $\delta$.

They should also work with tabular data, as banks normally use well-defined structured formats, and be publicly available for review by external parties. As public institutions, regulators should embrace clear, transparent, and accountable methods that can be discussed with stakeholders and mitigate reputational risks. Lastly, they should be compatible with user-supplied datasets such as banking microdata to be of practical use. Our work includes mechanisms from two different families so that the implementation is not constrained to a specific statistical approach.  

\subsection{Marginal-based inference and workload awareness}

Marginal-Based Inference (MBI) SDG approaches extract a set of marginal distributions from the original data, add noise, and use random sampling to produce synthetic data. As they use discrete marginals, inputs are required to be categorical. Recent benchmarking efforts show that MBI outperforms alternatives on tabular data applications \citep{tao2021benchmarking, bowen2019comparative}. Maximum-Spanning Trees (MST) are a differentially private mechanism in which marginals of interest are selected by constructing a graph, treating vertices as features and edges as the mutual information between them. After finding the MST \citep[see][]{mckenna2021winning}, Gaussian noise is added to the marginal distribution $f(d)$ to produce noisy marginals $M$ such that
\begin{eqnarray}
\label{gausnoise}
    M(d) = f(d) + \mathcal{N}(\mu, \sigma^{2}).
\end{eqnarray}
The mechanism then calculates the amount of noise that should be added to the sensitive dataset to satisfy the criteria, using the differential privacy parameters as 
\begin{eqnarray}\label{privacy}
    \sigma = \frac{\sqrt{\log(\frac{1}{\delta})} + \sqrt{\log(\frac{1}{\delta}) + \varepsilon}}{\varepsilon}.
\end{eqnarray}
Finally, a probabilistic graphical model is used to infer the new data distributions from the noisy marginals and generate synthetic samples \citep[for details, see][]{mckenna2021winning}. 

An alternative MBI mechanism is PAC-SYNTH, introduced by \citet{kim2021differentially}, which uses Private Accurate Combination (PAC). This approach relies on differentially private marginals to build synthetic data but uses an aggregate seeded synthesiser. Here, $k$-way noisy marginals, with $k$ as a user-set reporting length, are produced by extracting $k$-tuples from the dataset. Spurious tuples are controlled through a threshold parameter $\rho_k$ as 
\begin{eqnarray}\label{sp}
    \rho_k = \Delta_k^{1/2} \sigma_k \Phi^{-1} \left( 1 - \eta \min \left( 1, \frac{\mid S_{k-1} \mid}{\mid V_k \mid} \right) \right),
\end{eqnarray}
which, for a Gaussian CDF $\Phi$ and a maximum contribution parameter $\Delta_k$ (with a default value of 3), limits the probability that a spurious tuple may be included in the count by
\begin{eqnarray}\label{PAC}
    1 - \Phi(\frac{\rho_k}{\sigma_k \Delta_k^{1/2}}).
\end{eqnarray}
Here, $S_k$ represents a set of extracted $k$-tuples, $V_k$ is the set of $k$-tuples whose $(k-1)$ tuples belong to $S_k$, and $\eta \in (0, 1)$ denotes the proportion of spurious count in the noisy marginals. Noise is then added to the aggregated counts for DP compliance and to produce synthetic tabular data. 

A related class of approaches are workload-aware mechanisms such as the Adaptive and Iterative Mechanism (AIM) \citep{mckenna2022aim}. In some cases, the expected output, or workload, of synthetic data is known in advance, which can be used to optimise the privacy budget through prioritised marginals. AIM is closely related to MST but incorporates the workload parameter through a more sensitive quality score function to select marginals, adjusting the privacy spent on each marginal dynamically and selecting marginals of interest using the workload. 

\subsection{Generative adversarial networks}

Traditionally, GANs produce synthetic data with unstructured input such as images, but have been modified to work with tabular data and DP guarantees. The generative model transforms the input distribution to an output that approximates the data distribution, while the discriminative model estimates the probability of a sample being synthetic \citep{goodfellow2020generative}. 

In an SDG application, the generator model $\xi$ draws a random input vector $\textbf{x}$ from the probability distribution of the data $p(\textbf{x})$, and the discriminator model $\varphi$ evaluates the probability of it belonging to the original dataset by computing a value function $V(\xi,\varphi)$ as
\begin{eqnarray}
\label{GAN}
    \min_\xi \max_\varphi V(\xi,\varphi) = \mathbb{E}_{\textbf{x} \sim p(\textbf{x})}[\log(\varphi(\textbf{x})] + \mathbb{E}_{\textbf{z} \sim p(\textbf{z})}[\log(1 - \varphi(\xi(\textbf{z}))],
\end{eqnarray}
where $p(\textbf{z})$ is the noise distribution that is added to $\textbf{x}$. The loss error is calculated and backpropagated to the generator to adjust weights until the discriminator fails to distinguish. The underlying model can be trained with privacy costs; for example, differentially private GANs add noise to the gradient of the discriminator during training \citep{abadi2016deep, han2021differentially}. 

Likewise, a set of complementary techniques for tabular data, such as mode-specific normalisation, data imbalance, and training-by-sampling, are added by \citet{xu2019modeling} to create the Conditional Tabular GAN (CTGAN). One GAN mechanism we select is the Differentially Private CTGAN (DPCTGAN) as an adaptation of the CTGAN framework, which uses noise and gradient clipping techniques to provide DP guarantees. Here, the Wasserstein distance is used to approximate the distance between probability distributions in the loss function as 
\begin{eqnarray}
\label{DPGAN}
    \min_\xi \max_{w\in W} \mathbb{E}_{\textbf{x}\sim p(\textbf{x})}[f_w(\textbf{x})]-\mathbb{E}_{\textbf{z}\sim p(\textbf{z})}[f_w(\xi(\textbf{z}))].
\end{eqnarray}
PATECTGAN, on the other hand, supports DP through several models trained on disjoint datasets $D_i$, so that no single model uses all of the data and cannot access an individual teacher $T_i$ \citep{papernot2016semi}.\citet{jordon2018pate} use PATE to modify the training procedure of the discriminator model. A new input feature vector $\textbf{x}$ is aggregated with noise for $m$ possible classes using 
\begin{eqnarray}
\label{PATEGAN}
    n_j(\textbf{x}) = | \{ T_i: T_i(\textbf{x}) = j \} |,
\end{eqnarray}
where $ j \in \{1, \dots, m\}$ and $n_j(\textbf{x})$ is the number of teachers that output class $j$ for the input vector. The corresponding output $\Omega$ for the input vector $\textbf{x}$ is defined as 
\begin{eqnarray}
\label{PATEGAN2}
    \Omega(\textbf{x}) = \arg\max_{j} (n_j(\textbf{x}) + Y_j),       
\end{eqnarray}
where $Y_j$ are the Laplace noise variables. We build these mechanisms using OpenDP's SmartNoise\footnote{\url{https://docs.smartnoise.org}} SDK package and use default configurations, except for GAN mechanisms, where we reduce the epoch limit from $300$ to $100$ due to no discernible improvement during trial runs. SDG mechanisms commonly allocate a privacy budget for integrated pre-processing transformations, but in this case, the datasets were transformed outside of the mechanisms. The entire pre-processed dataset is used to train and configure synthetic data with the same dimensions as the original microdata. 

\subsection{Domain pre-processing and applications} 
\label{sec:domainpre}

Domain pre-processing covers the necessary encoding and decoding techniques dictated by the limitations of mixed data types. MBI mechanisms only work on categorical datasets due to using discrete marginals, while GANs require numerical features for gradient descent updates. Some numerical features in CBP's banking microdata have very large domains because of the local currency (in the order of $10^{11}$), or present non-normal or multi-modal distributions.

Our work includes two categorical pre-processing strategies (PS). One uses the current data taxonomy of the CBP, mainly through domain transformations partitioning numeric features with the bank's cut-offs to assess the alignment with current statistical disclosure. The second  incorporates data-driven transformations based on, for example, binning and logarithmic transformations.

Bin widths could be non-uniform but maintain the same number of data points across bins. Observations in a feature are sorted and divided into $k$ groups for numeric features such as term and capital, with minimum and maximum values set as bin widths, to mitigate class imbalance. We also use $k$-means binning with features such as credit card debt to represent underlying multi-modal and skewed distributions. Here, bin widths are defined by treating observations as clusters, using one-dimensional $k$-means to group observations $x$ and $y$ into $k$ sets $S$ as
\begin{eqnarray}
\label{k-means}
    \argmin_S \sum_{i=1}^k \frac{1}{\mid S_i \mid} \sum_{x,y \in S_i} \parallel x - y \parallel^2.
\end{eqnarray}
Numeric features with a very large domain, such as debt and capital, are log-transformed beforehand. Additionally, an optional post-processing module is implemented to convert the data back (numerical or categorical). We use a Gaussian Kernel Density Estimator (KDE) to smooth the histogram of each feature in the original banking microdata to avoid individual information leakage. The Gaussian KDE is sampled using an array of values from the minimum and maximum value of each feature, at a random interval. The KDE values are then used as probabilities and sampled with replacement. These post-processed datasets are part of the published data.

Existing SDG benchmarking is commonly centred on general applicability. However, financial regulators already release summary statistics of sensitive data, and an evaluation through domain-specific information products allows them to gauge utility and seek public and expert feedback. We thus provide an additional contribution through a utility-first approach, where synthetic data is designed with a specific application in mind to determine if synthetic banking microdata could be used to replicate existing statistical disclosure can expand granularity.

In this work, we develop three domain-specific applications to evaluate synthetic banking microdata utility, which relate to core processes at the Central Bank of Paraguay: financial inclusion, monetary policy, and credit risk management. The corresponding datasets\footnote{The URL to the dataset will be included here in the final version for publication.} are made publicly available through the Zenodo open repository under the European OpenAIRE programme.

\section{Use Cases}
\label{sec:use_cases}

\subsection{Application 1: A financial usage index}

Financial Inclusion (FI) has been recognised as a main pillar of the global development agenda at 2010's G20 Summit. Many regulators have placed it under their umbrella of responsibilities for more inclusive economic development, poverty reduction, economic growth, and financial stability \citep{omar2020does}. International and widely accepted FI standards remains elusive, with research focussing on cross-market inclusion and effective promotion. 

The measurement of FI was advanced by the World Bank's Global Findex Database (GFD) in 2011, using survey data on 123 countries for comparable indicators \citep{klapper2022little}. Implementations of a multidimensional index include dimensionality reductions like two-stage Principal Component Analysis (PCA) and factor analysis for data-driven model parameters \citep{camara2014measuring}. GFD indicators are compiled from governmental and other official sources, but less-developed economies historically face challenges in data collection. Microdata offers a fine-grained view of features like age and gender, which are previously inaccessible inclusion indicators.  

Our first objective is to evaluate whether such data has benefits relative to panel data such as the GFD. Microdata is used to replicate the latent variable FI index linearly determined by \citet{camara2014measuring} with a usage ($B^u$), barrier ($B^b$), and access ($B^a$) component. Banking microdata relates to the use of financial services and products in a country $i$ in three main indicators,
\begin{eqnarray}
\label{eq:usage_component_origi}
    B_i^u = \psi_1 \alpha_i + \psi_2 \beta_i + \psi_3 \gamma_i + \nu_i , 
\end{eqnarray}
where $\alpha_i$ is the percentage of the population with a bank account (or mobile banking services and credit cards), $\beta_i$ the one that saves using a bank account, $\gamma_i$ the one with a loan from a formal financial institution, and $\nu$ is the variation due to error. $\psi$ and $B_i^u$ indicate unobserved variables that need to be defined. We select survey indicators $o$ that are potentially replaceable with banking microdata to reproduce a usage component $B_{o}^u$ using the PCA transformation described in \citet{camara2014measuring}. The microdata is used to generate a dataset replicating the three usage indicators with demographic characteristics listed in Table ~\ref{tab:usecase1_features} from 2017--2023.  

\begin{table}[!htb]
\caption{Financial inclusion microdata features with post-processing strategies (PS). The latter are provided for both those used by the Central Bank of Paraguay (CBP) and for a data-driven (DD) approach.}
\begin{footnotesize}
\begin{center}
\begin{tabular}{ c | l r}
\toprule
\textbf{PS}& \textbf{Feature}  & \textbf{Binning (levels)}\\
\midrule
\multirow{10}{*}{CBP}
&Period&${[2017--2023]}$ (6)\\
\cline{2-3}
&Age&${[<25,35,45,55,65,75,110]}$ (7)\\
\cline{2-3}
&Gender&${[M,F]}$ (2)\\
\cline{2-3}
&nCCards&${[0,1,2,3,>3]}$ (5)\\
\cline{2-3}
&hasCollateral&${[0,\geq(1)]}$ (2)\\
\cline{2-3}
&nLoans&${[0,1,2,3,>3]}$ (5)\\
\cline{2-3}
&loanMaxDuration&${[0,<400,740,1100,1850,>1850]}$ (6)\\
\cline{2-3}
&nFI&${[1,2,3,>3]}$ (4)\\
\cline{2-3}
&nNZS&${[0,1,2,3,>3]}$ (5)\\
\cline{2-3}
&nSavings&${[0,1,2,3,>3]}$ (5)\\
\hline
\multirow{10}{*}{DD}
&Period&${[2017--2023]}$ (6)\\
\cline{2-3}
&Age&uniform intervals (17)\\
\cline{2-3}
&Gender&{[M,F]} (2)\\
\cline{2-3}
&nCCards&k-means (5)\\
\cline{2-3}
&hasCollateral&${[0,\geq(1)]}$ (2)\\
\cline{2-3}
&nLoans&k-means (5)\\
\cline{2-3}
&loanMaxDuration&k-means (4)\\
\cline{2-3}
&nFI&k-means (7)\\
\cline{2-3}
&nNZS&equal-frequency (4)\\
\cline{2-3}
&nSavings&k-means (5)\\
\bottomrule
\end{tabular}
\label{tab:usecase1_features}
\end{center}
\end{footnotesize}
\end{table}

One challenge of such microdata is the lack of information on the unbanked population. However, other government sources such as Paraguay's personal identity registry, can be used to add the missing values of the unbanked population with respect to age and gender. Through this combination, we recreate the percentage domain used in \citet{camara2014measuring} to calculate Paraguay's usage indicators $\alpha_o$, $\beta_o$, and $\gamma_o$. We then generate data with the chosen SDG mechanisms to recompute usage indicators and evaluate their utility under strong privacy requirements. First, we treat the banking microdata using the CBP's pre-processing strategies covered in Section \ref{sec:domainpre}, where the usage component built with synthetic data $B_s^u$ as in Equation~\ref{eq:usage_component_origi}.

Synthetic indicators are produced for each individual such that $\alpha_s$ and $\beta_s$ are the number of financial institutions and the number of savings accounts, respectively, while $\gamma_s$ denotes the number of loan operations (see Table ~\ref{tab:usecase1_features}). Age is categorised into seven bins used in other demographic reporting. We add the unbanked population using `None' and calculate $B_s^u$.

At this point, the error between each mechanism's $B_s^u$ and $B_o^u$ values indicates the proximity between usage components. In this application, we compute the maximum absolute difference $\tau$ between $B_{s,m}^{u}$ and $B_o^u$, where $m$ denotes the SDG mechanism. The final step is to extend the granularity of the usage FI index with the data-driven PS. We use the attributes of each feature to approximate an indicator $B_{s,l}^u$, where $l$ separates the banked population into low, medium, and high usage. The highest indicator level defines financial access, savings, and loans condition $B_{s,l}^u$.

\subsection{Application 2: Term deposit yield curves}
\label{sec:term_deposit}

One of the responsibilities that central banks share is price stability. Regulators closely monitor the money supply as it flows through the financial system. Particular attention is paid to institutions that act as maturity transformation agents, matching short-term savings with long-term financing demanded by the market \citep{hakenes2019deposit}. The value of money is represented by a yield curve showing interest rates required for a term instrument at a range of maturities. Central banks adjust these rates so banks adjust their lending and deposits rates, ultimately aligning the money supply with their monetary policy \citep{chang2010lending}. 

The effectiveness of this interaction is important to regulators, as yield curves produced with government debt can serve as an indicator of future economic activity, arguing that liquidity demand is connected to economic expectations \citep{gurkaynak2007us}. However, in some developing economies such as Paraguay, term deposits remain a significant part of liquidity funding. In 2021, the nominal volume of bonds negotiated through Paraguay's stock exchange totalled approximately 1.2 billion USD\footnote{https://siv.bcp.gov.py/\#menu1}, while the outstanding volume of government bonds was 760 million USD\footnote{https://economia.gov.py/index.php/financiamiento} and term deposit instruments were close to 7 billion USD\footnote{https://www.bcp.gov.py/userfiles/files/1\_BOLB\%20122021\%281\%29.xls} (approximately 18\% of GDP). 

Constructing a yield curve is not trivial, and might not only depend on model selection but also the underlying optimisation method \citep{manousopoulos2009comparison}. The CBP currently uses microdata to report aggregated statistics on term deposits. Each month, the average interest rate (weighted with respect to capital), the total capital, and the count of issued term deposit operations are disclosed using pre-defined term bins.

Unfortunately, there are several limitations. While only the weighted average of two independent features, capital and nominal interest rate, is disclosed, other statistics such as the median and quantiles may be equally relevant as numerical banking features usually have skewed distributions \citep{adcock2015skewed}. Weighting by capital could also distort the expected average rates, which is further magnified by non-standardised and unscaled. This is particularly detrimental to potential savers and investors interested in the market rate of a specific capital range. 

Our application implements a more granular term deposit yield curve. Relevant features are selected from CBP microdata from 2019--2023 and include information from the term deposit curve, with numeric features transformed using the CBP's aforementioned PS. The 28 term bins used in the yield curve are maintained, but capital is binned using a scale of multiples of the deposit insurance limit shown in Section \ref{sec:domainpre}. Similarly, the nominal interest rate is discretised using equal-width bins of 0.5\%, ranging from 0\% to 14\%, as shown in Table ~\ref{tab:usecase2_features}. 

\begin{table}[!htb]
\caption{Term deposit yield curve microdata features with post-processing strategies (PS). The latter are provided for both those used by the Central Bank of Paraguay (CBP) and for a data-driven (DD) approach. Here, $L$ denotes the deposit insurance limit, currently 200 million PYG.}
\begin{footnotesize}
\begin{center}
\begin{tabular}{ c | l r}
\toprule
\textbf{PS strategy}& \textbf{Feature}  & \textbf{Binning (levels)}\\
\midrule
\multirow{6}{*}{CBP}
&typeFI&[`Bank', `Nonbank'] (2)\\
\cline{2-3}
&Period&[12/2019 - 12/2023] (5)\\
\cline{2-3}
&Currency&[`USD', `PYG'] (2)\\
\cline{2-3}
&Capital&$L(2^n)$ for $n \in \{-1,0,...,7\}$ (9)\\
\cline{2-3}
&Term& [$<30,30,60,...,>3600$] (28)\\
\cline{2-3}
&Interest Rate& [0,0.5,...15] (16)\\
\hline
\multirow{6}{*}{Data-driven}
&Type FI&[`Bank', `Nonbank'] (2)\\
\cline{2-3}
&Period&[12/2019 - 12/2023] (5)\\
\cline{2-3}
&Currency&[`USD', `PYG'] (2)\\
\cline{2-3}
&Capital&equal-frequency (5)\\
\cline{2-3}
&Term&equal-frequency (5)\\
\cline{2-3}
&Interest Rate&k-means (5)\\
\bottomrule
\end{tabular}
\label{tab:usecase2_features}
\end{center}
\end{footnotesize}
\end{table}

We then use the pre-processed dataset and the selected SDG mechanism to produce synthetic data. The capital and nominal interest rate features are converted back to numerical values using the left-most bin edge to compute the weighted interest rate, and a new yield curve $Y_{m}$ is produced for each mechanism $m$. The maximum RMSE for all term bins $n$ in the set of periods $P$ is calculated as the error $\upsilon_{m}$ between yield curves for synthetic and original microdata $Y_{o}$,
\begin{eqnarray}
\label{eq:RMSE}
    \upsilon_{m} = \max_{p \in P} \sqrt{\frac{1}{n} \sum_{i=1}^{n} (\hat{Y}_{i,m,p} - Y_{i,o,p})^2}.
\end{eqnarray}
The second objective is to evaluate whether synthetic data can lead to a more granular yield curve. In this case, the respective dataset uses the data-driven PS defined in Table ~\ref{tab:usecase2_features} for capital, term, and nominal interest rate. An issue with the CBP's current yield curve is that a line plot representation does not convey the general trend effectively. To account for this, we use Locally Weighted Scatterplot Smoothing (LOWESS) to determine a regression fit and assess whether a more general trend can be described directly \citep{cleveland1981lowess}.

\subsection{Application 3: Credit card transition matrices} 
\label{sec:credit_Card}

Our third application relates to credit risk management, as the ability for systemic and structural risk to spread between institutions is a concern for regulators. The Federal Reserve Economic Data portal shows that credit card debt serves as a proxy for household debt and consumption, and delinquency is used to track economic sentiment and consumer spending \citep{fulford2020revolving}. In operational research, credit card modelling focuses on applications such as forecasting credit card usage \citep{alfonso-sanchez2024optimizing}, scoring models \citep{silva2022class}, and estimating loss such as exposure at default \citep{leow2016new, wattanawongwan2023modelling}.

The CBP presently uses banking microdata to disclose account-level reports on the total debt and number of credit cards in the banking system, but no data is provided on delinquency or non-performing loan status. We therefore use synthetic banking microdata to enhance the credit card information available in risk reporting. As the first step, the CBP's reporting is replicated using synthetic data for the years 2020 and 2021. 

Most features in this dataset are explicit fields that financial institutions report monthly, such as debt levels and delinquency (in days). Domain definitions such as delinquency days are especially sensitive due to their importance to firm-level financial oversight such as the supervision of solvency, capital adequacy requirements, and risk management of individual financial institutions. The dataset is transformed using the CBP's PS as listed in Table ~\ref{tab:usecase3_features}. 

\begin{table}[!htb]
\caption{Credit card transition matrix microdata features with post-processing strategies (PS). The latter are provided for both those used by the Central Bank of Paraguay (CBP) and for a data-driven (DD) approach. Here, $W$ denotes 2020's minimum monthly wage, which is approximately 2.1 million PYG.}
\begin{footnotesize}
\begin{center}
\begin{tabular}{ c | l r}
\toprule
\textbf{PS strategy}& \textbf{Feature}  & \textbf{Binning (levels)}\\
\midrule
\multirow{6}{*}{CBP}
&Gender&${[M,F]} (2)$\\
\cline{2-3}
&Age2020&${[<25,35,45,55,65,75,110]} (7)$\\
\cline{2-3}
&Debt2020&$W(2^n)$ for $n \in \{-1,0,...,7\}$(8)\\
\cline{2-3}
&Debt2021&$W(2^n)$ for $n \in \{-1,0,...,7\}$(8)\\
\cline{2-3}
&Delinquency2020&[$<61,91,151,181,271,>271$](6)\\
\cline{2-3}
&Delinquency2020&[$<61,91,151,181,271,>271$](6)\\
\cline{2-3}
\hline
\multirow{6}{*}{Data-driven}
&Gender&${[M,F]} (2)$\\
\cline{2-3}
&Age2020&k-means (6)\\
\cline{2-3}
&Debt2020&k-means (7)\\
\cline{2-3}
&Debt2021&k-means (7)\\
\cline{2-3}
&Delinquency2020&k-means (6)\\
\cline{2-3}
&Delinquency2020&k-means (6)\\
\cline{2-3}
\bottomrule
\end{tabular}
\label{tab:usecase3_features}
\end{center}
\end{footnotesize}
\end{table}

The pre-processed dataset is used to generate synthetic data with the chosen SDG mechanisms. In the data-driven PS, we discretise numeric features through $k$-means transformations, as credit management actions such as selling non-performing loans result in non-normal and multi-modal distributions. We then build credit card transition matrices using the microdata generated by both pre-processing strategies. Such matrices have been used to analyse credit card default levels \citep{leow2014intensity, stepankova2021bank}. They are employed in credit risk modelling to reflect transitions between delinquency levels through a square matrix $C$, for the probabilities of state $j$ transitioning to $i$, or $P(j \mid i) = P_{i,j}$, using the data counts between two periods. This is also extended to transitions between debt levels, which may be useful to evaluate exposure at default. 

We evaluate the utility error between synthetic and original microdata using the relative delinquency and generated transition matrices. For each mechanism $m$, the transition matrix $A_m$ and the same based on original microdata ($A_o$) are used to compute the difference matrix $M_m= A_m - A_o$, for delinquency ($D_{\mathrm{del},m}$) and total debt ($D_{\mathrm{deb},m}$). The Frobenius norm is computed to use the difference magnitude as a metric of relative error, as demonstrated by \citet{malioutov2014convex}. Here, $\kappa$ is the difference between elements in the ($n \times n$) transition matrix generated with synthetic and original data, so that the Frobenius norm is
\begin{equation}
\label{eq:fro}
    \parallel M_{m} \parallel_F = \sqrt{\sum_{i = 1}^n \sum_{j = i}^n \mid \kappa_{i,j} \mid^2}. 
\end{equation}
Like other utility-first approaches, for this initial exploration, we set conservative parameters for the built-in DP guarantees of mechanisms. Setting $\varepsilon = 1$ and $\delta = 10^{-10}$ provides strong privacy guarantees \citep[see, for example,][]{cohen2022private, dwork2010differential}. 

\section{Results}

\subsection{Data preparation and model configuration}

We first need to transform the raw data into a usable format specific to each domain application. Financial inclusion, for example, requires loan and savings information for every individual, using the entire banking dataset for approximately 3.7 million individuals and each time period. Likewise, features such as the type of financial instrument have hundred of levels that are mapped into three types of operations: credit cards, savings, and loans.

Some features include past reporting errors because no effective data validation controls were implemented in the past, and reporting mechanics and standards have changed over time. At the beginning of the savings microdata collection, some institutions report interest rates of short-term savings instruments using a monthly instead of an annualised interest rate. External data sources were previously also not centralised, resulting in data entry errors. 

Demographic datasets were not automatically retrieved from the national registry but manually reported by each financial institution, leading to errors in assigned gender, implausible ages, and missing features. To account for this, we design custom controls to exclude rows with inconsistencies. This process also demonstrates that data validation and quality remain a definite concern for financial regulators, even with highly structured regulatory banking microdata.  

Each application requires two versions of its microdata; one using the CBP's and one using the data-driven PS (see Tables~\ref{tab:usecase1_features}, ~\ref{tab:usecase2_features}, and ~\ref{tab:usecase3_features}). As these transformations take place outside of the mechanisms, a privacy budget for integrated pre-processing is not necessary. While we keep default configurations for mechanisms across domain applications, the epoch limit for GANs is reduced from $300$ to $100$, as no discernible improvement is reached after that threshold during trial runs.

\subsection{Evaluation of the financial usage index}

In Table ~\ref{tab:usage_components}, we replicate Paraguay's usage component $B_o^u$ with microdata from 2017--2023, as the CBP's current data collection began in 2016. The GFD features data points for 2011, 2017 and 2021, which are used to recalculate the usage component using the PCA transformation in \citet{camara2014measuring}. One explanation for the difference in the increase is the introduction of `simplified accounts' in 2013, which are basic savings accounts with minimum identity verification requirements. Regulation mandates that digital payment providers open these at banking institutions on behalf of clients, leading to a stark increase in the number of accounts. We transform the original dataset using the CBP's pre-processing strategy and generate synthetic data with each mechanism to compute the usage component $B_{s}^u$ for each period.

\begin{table}[!htb]
\caption{Paraguay's usage component $B_o^u$ as replicated with banking microdata for 2017--2023 and the usage component $B_\mathrm{PY}^u$ calculated with survey data from the 2011 Global Findex Database.}
\begin{footnotesize}
\begin{center}
\begin{tabular}{ l r r }
\toprule
\multicolumn{1}{c}{\textbf{Year}} & \multicolumn{1}{c}{\textbf{Microdata}} & \multicolumn{1}{c}{\textbf{Survey}}\\
\midrule
$2011$&-&$0.33$\\
$\dots$&$\dots$&$\dots$\\
$2017$&$0.57$&$0.48$\\
$2018$&$0.61$&-\\
$2019$&$0.60$&-\\
$2020$&$0.56$&-\\
$2021$&$0.57$&$0.44$\\
$2022$&$0.59$&-\\
$2023$&$0.69$&-\\
\bottomrule
\end{tabular}
\label{tab:usage_components}
\end{center}
\end{footnotesize}
\end{table}

Error metrics calculated with respect to the original usage component are low and consistent across periods for MBI mechanisms. As a major reason for using banking microdata is improved granularity, we build a FI index with demographic variables. Metrics are recalculated for each demographic group to evaluate how error propagates into higher granularities.

\begin{table}[!htb]
\caption{Maximum absolute difference ($\tau$) between usage components $B_{o}^u$ and $B_{s}^u$, and its distribution among age bands and gender, for periods (2017--2023). $B_{s}^u$ is pre-transformed with the CBP's PS.}
\begin{footnotesize}
\begin{center}
\begin{tabular}{ l l c r r r r r}
\toprule
\multicolumn{2}{c}{\textbf{Demographic}} & \multicolumn{1}{c}{} & \multicolumn{5}{c}{\textbf{Mechanism}} \\
\cline{1-2} \cline{4-8}
\multicolumn{1}{c}{\textbf{Age Group}} & \multicolumn{1}{c}{\textbf{Gender}} & \multicolumn{1}{c}{} & \multicolumn{1}{c}{\textbf{MST}} & \multicolumn{1}{c}{\textbf{AIM}}& \multicolumn{1}{c}{\textbf{PAC}}& \multicolumn{1}{c}{\textbf{DPCTGAN}}& \multicolumn{1}{c}{\textbf{PATECTGAN}}\\
\midrule
\multirow{2}{*}{$\mathrm{<25}$} 
&M& &$0.019$&$0.014$&$0.024$&$0.405$&$0.601$\\
&F& &$0.023$&$0.014$&$0.060$&$0.347$&$0.574$\\
\hline
\multirow{2}{*}{$\mathrm{25-35}$} 
&M& &$0.008$&$0.004$&$0.114$&$0.812$&$0.154$\\
&F& &$0.014$&$0.005$&$0.042$&$0.593$&$0.281$\\
\hline
\multirow{2}{*}{$\mathrm{36-45}$}
&M& &$0.008$&$0.002$&$0.119$&$0.914$&$0.495$\\
&F& &$0.021$&$0.003$&$0.040$&$0.802$&$0.287$\\
\hline
\multirow{2}{*}{$\mathrm{46-55}$}
&M& &$0.018$&$0.008$&$0.079$&$0.353$&$0.979$\\
&F& &$0.017$&$0.009$&$0.014$&$0.573$&$0.378$\\
\hline
\multirow{2}{*}{$\mathrm{56-65}$}
&M& &$0.023$&$0.008$&$0.067$&$0.458$&$1.007$\\
&F& &$0.017$&$0.008$&$0.070$&$0.672$&$0.907$\\
\hline
\multirow{2}{*}{$\mathrm{66-75}$}
&M& &$0.134$&$0.011$&$0.260$&$0.685$&$1.849$\\
&F& &$0.070$&$0.012$&$0.321$&$0.550$&$1.499$\\
\hline
\multirow{2}{*}{$\mathrm{76+}$}
&M& &$0.154$&$0.013$&$0.728$&$0.650$&$4.723$\\
&F& &$0.091$&$0.013$&$0.534$&$0.520$&$2.231$\\
\bottomrule
\end{tabular}
\label{tab:app1_absolute_difference}
\end{center}
\end{footnotesize}
\end{table}

Table ~\ref{tab:app1_absolute_difference} shows that the absolute difference of $B_{s}^u$ is not distributed uniformly across age and gender for all mechanisms. Some mechanisms are promising for general metrics such as a financial index, but need to be treated with caution, as PAC could result in biased distributions for some age groups due to suppressing aggregated counts below a certain threshold, while MST and AIM maintain low error rates at general and more granular levels. Table ~\ref{tab:top4_demo_usingAIM} uses synthetic data generated by AIM to show the change in the top four usage components for gender and age groups.

\begin{table}[!htb]
\caption{Top four usage components $B_{s}^u$ for 2017 and 2023 using AIM-generated synthetic banking microdata.}
\begin{footnotesize}
\begin{center}
\begin{tabular}{ l l r c l l r}
\toprule
\multicolumn{3}{c}{\textbf{2017}} & \multicolumn{1}{c}{} & \multicolumn{3}{c}{\textbf{2023}}\\
\cline{1-3} \cline{5-7}
\textbf{Age group} & \textbf{Gender} & \textbf{$\mathrm{B_{s}^u}$} & & \textbf{Age group} & \textbf{Gender} & \textbf{$\mathrm{B_{s}^u}$}\\
\midrule
36-45&M&0.80& &36-45&M&0.91\\
25-35&M&0.71& &25-35&M&0.88\\
46-55&M&0.71& &25-35&F&0.80\\
56-65&M&0.66& &36-45&F&0.80\\
\bottomrule
\end{tabular}
\label{tab:top4_demo_usingAIM}
\end{center}
\end{footnotesize}
\end{table}

\begin{figure}[!htb]
    \centering
    \includegraphics[width=\textwidth]{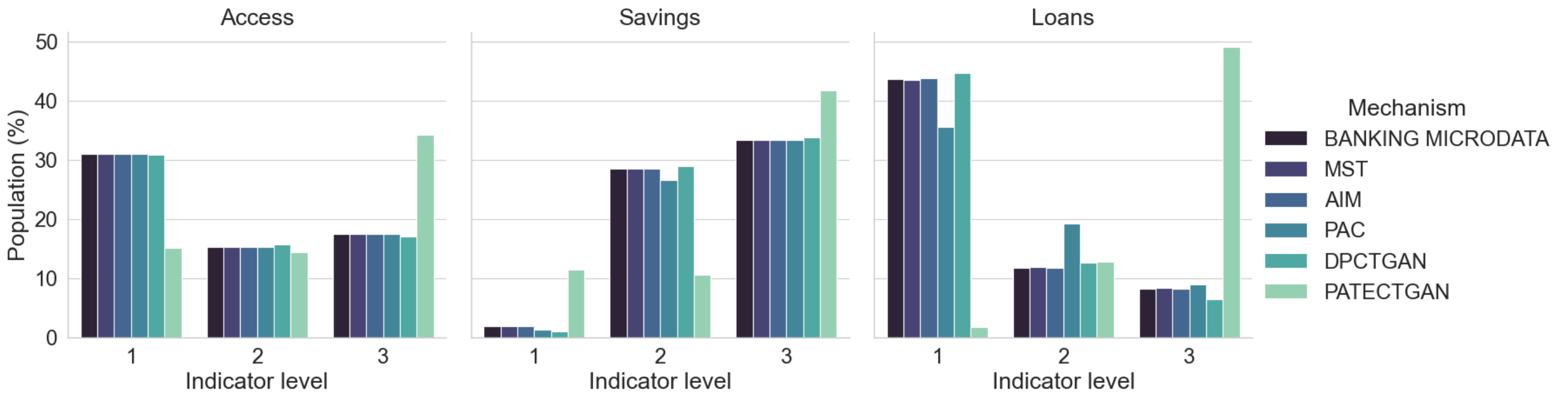}
    \caption{Paraguay's banked population by usage indicators $B_{s,l}^u$ with levels low (1), medium (2), and high (3). Usage indicators are calculated with (original) banking and synthetic microdata from 2023.}
    \label{fig:figure_1}
\end{figure}

We implement the same for the data-driven PS, with usage indicators in Table~\ref{tab:usecase1_features} separating the banked population into low, medium, and high usage. 
MBI mechanisms again outperform GANs in approximating the original relative population of each usage level $B_{s,l}^u$. The sub-components' level distributions in Figure \ref{fig:figure_1} provide insights into differences of financial inclusion penetration behaviour for savings and loans products. In terms of savings, most of the banked population is at a medium-high usage level, which could be explained by simplified accounts, with a limit of two per individual. In contrast, most of the banked population in the loan sub-component is in the low indicator level, mirroring the higher difficulty of obtaining loans. 

\subsection{Comparison of term deposit yield curves}

The application to term deposit yield curves aims to improve upon a current information product disclosed by the CBP. We replicate the curves with banking microdata as a benchmark, group the data with the 28 bins defined by the CBP, and calculate the weighted interest rate (with respect to capital) for the end of 2019 to 2023. The discontinuity due to missing data in term bins presents a challenge, as it can result in counterintuitive representations of curves. 

\begin{figure}[!htb]
    \centering
    \includegraphics[width=\textwidth]{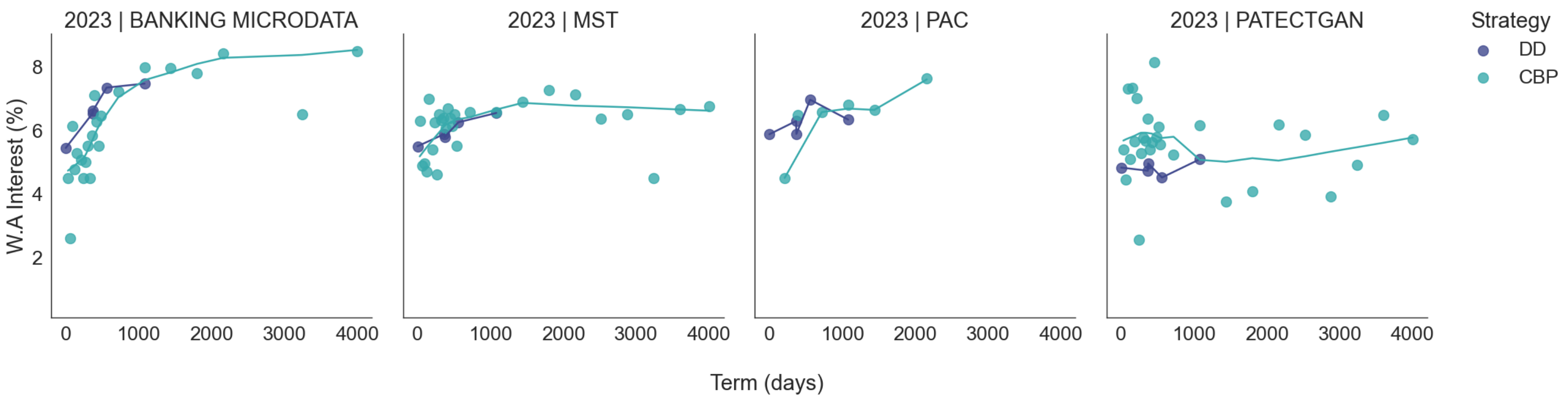}
    \caption{Yield curves produced with synthetic term deposit microdata using MST, PATECTGAN, and PAC, for the Central Bank of Paraguay's (CBP) and data-driven (DD) pre-processing. Each subplot includes the scatter points of the calculated weighted average interest rate for term bins and a LOWESS curve.}
    \label{fig:figure_2}
\end{figure}

We first use the dataset with the CBP's PS to generate synthetic data and compute the weighted interest rate average following Section \ref{sec:term_deposit}. Figure \ref{fig:figure_2} shows that the time series produced by SDG mechanisms fail to preserve the structure or a low error bound to the original yield curves, with missing data and low counts in certain term bins able to impact SDG performance. PAC only considers bins with counts above a certain threshold and suppresses the rest, producing only four data points for each period. The effect of low data counts is further amplified by the strict DP parameters, which make mechanisms suppress or add too much noise to the frequency counts.   

Even the top-performing mechanisms in the previous application (MST and AIM) provide limited results, with one likely source of error being the post-processing information loss resulting from the conversion of interest rates and capital bins to numerical values, as each instrument is weighted equally. However, capital influences both the interest rate and the term maturity, meaning some financial institutions unlock higher interest rates with account-specific variables such as capital and maturity thresholds \citep{bikker2018determinants}.

\begin{figure}
    \centering
    \includegraphics[width=\textwidth]{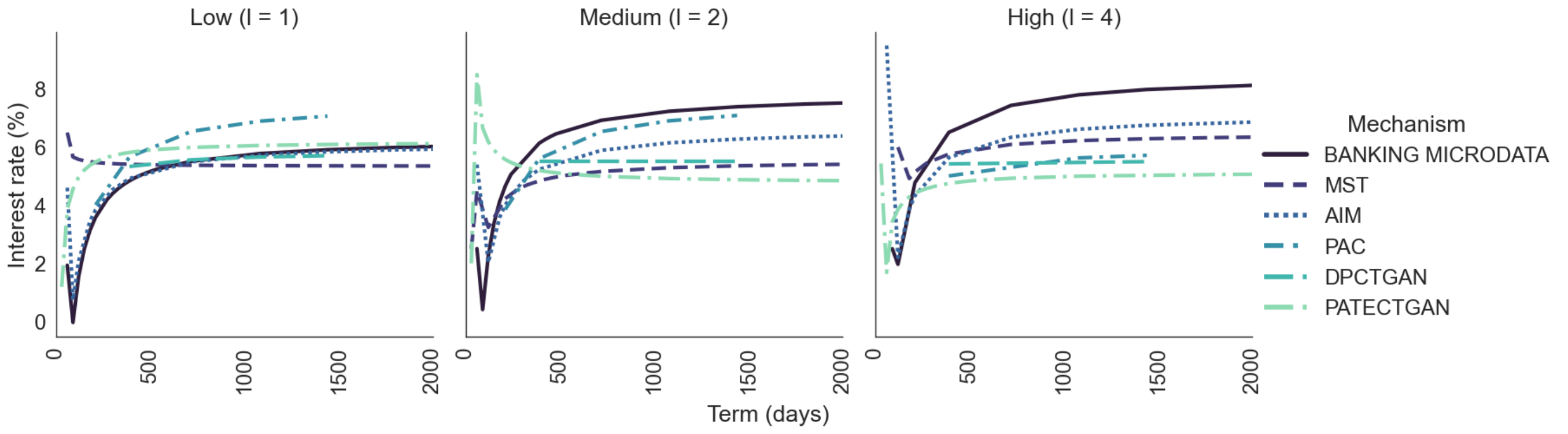}
    \caption{Nelson-Siegel-Svensson curves calculated with synthetic data for each mechanism, with the CBP's pre-processing strategy and 2023 term deposit data.}
    \label{fig:figure_3}
\end{figure}

In Figure~\ref{fig:figure_3}, despite the large error in term bins, synthetic microdata remains somewhat useful in reflecting how the weighted average interest rate changes for low levels of capital when using a parametric model such as Nelson-Siegel-Svensson \citep{nelson1987parsimonious, svensson1994estimating}. Table ~\ref{tab:app2_RMSEdifference} shows the difference between synthetic and original yield curves using the RMSE (see Equation \ref{eq:RMSE}). 
Here, all mechanisms average relatively high error with regard to the original WAI and TC domain (1\% - 10\% and 47 billion PYG, respectively).

\begin{table}[!htb]
\caption{RMSE for weighted average interest (WAI) rate and total capital (TC) for PYG and USD using synthetic microdata, with the Central Bank of Paraguay's (CBP) and the data-driven (DD) pre-processing strategies. The maximum RMSE value of all periods and type of financial institution is shown.}
\begin{footnotesize}
\begin{center}
\begin{tabular}{ l l l c r r r r r}
\toprule
\multicolumn{1}{c}{\textbf{}} & \multicolumn{1}{c}{\textbf{}} & \multicolumn{1}{c}{\textbf{}} & \multicolumn{1}{c}{\textbf{}} & \multicolumn{5}{c}{\textbf{Mechanism}} \\
\cline{5-9}
\multicolumn{1}{c}{\textbf{Type}} & \multicolumn{1}{c}{\textbf{Currency}} & \multicolumn{1}{c}{\textbf{Strategy}}& \multicolumn{1}{c}{\textbf{}} & \multicolumn{1}{c}{\textbf{MST}} & \multicolumn{1}{c}{\textbf{AIM}}& \multicolumn{1}{c}{\textbf{PAC}}& \multicolumn{1}{c}{\textbf{DPCTGAN}}& \multicolumn{1}{c}{\textbf{PATECTGAN}}\\
\midrule
\multirow{4}{*}{WAI} & \multirow{2}{*}{PYG} 
  &CBP& &$5.74$&$5.95$&$6.60$&$7.48$&$4.98$\\
& &DD& &$1.24$&$0.93$&$1.95$&$5.20$&$3.17$\\
&\multirow{2}{*}{USD}
  &CBP& &$4.26$&$5.09$&$4.06$&$5.20$&$5.39$\\
& &DD& &$1.79$&$1.49$&$2.15$&$4.69$&$4.21$\\
\hline
\multirow{4}{*}{TC} & \multirow{2}{*}{PYG} 
  &CBP& &$2.1\mathrm{e}{10}$&$2.2\mathrm{e}{10}$&$2.4\mathrm{e}{10}$&$3.3\mathrm{e}{11}$&$1.0\mathrm{e}{11}$\\
& &DD& &$2.6\mathrm{e}{10}$&$1.1\mathrm{e}{10}$&$3.5\mathrm{e}{10}$&$5.3\mathrm{e}{10}$&$5.4\mathrm{e}{10}$\\
&\multirow{2}{*}{USD}
  &CBP& &$5.7\mathrm{e}{10}$&$5.9\mathrm{e}{10}$&$8.0\mathrm{e}{10}$&$9.8\mathrm{e}{10}$&$9.8\mathrm{e}{10}$\\
& &DD& &$4.8\mathrm{e}{10}$&$2.0\mathrm{e}{10}$&$7.2\mathrm{e}{10}$&$2.0\mathrm{e}{11}$&$8.8\mathrm{e}{10}$\\
\bottomrule
\end{tabular}
\label{tab:app2_RMSEdifference}
\end{center}
\end{footnotesize}
\end{table}

Next, we produce yield curves using the data-driven PS defined in Table ~\ref{tab:usecase2_features} to check whether the selection of bins reduces this error. The weighted interest rates computed with the data-driven PS are closer to the expected output, especially with MST and AIM (see Table ~\ref{tab:app2_RMSEdifference}). However, the term's domain produced with the data-driven PS has been reduced significantly (1,000 vs. 5,000 days), an expected consequence of equal-frequency compressing outliers into a single bin. Our results indicate that an improved term deposit yield curve could be obtained if the PS strategy represented the underlying distribution more effectively. More generally, this application shows that pre-processing can be tuned for potential domain-applications until an optimum trade-off between SDG performance and utility is reached. 

\subsection{Credit card transition matrix outcomes}

\begin{figure}[!htb]
    \centering
    \includegraphics[width=\textwidth]{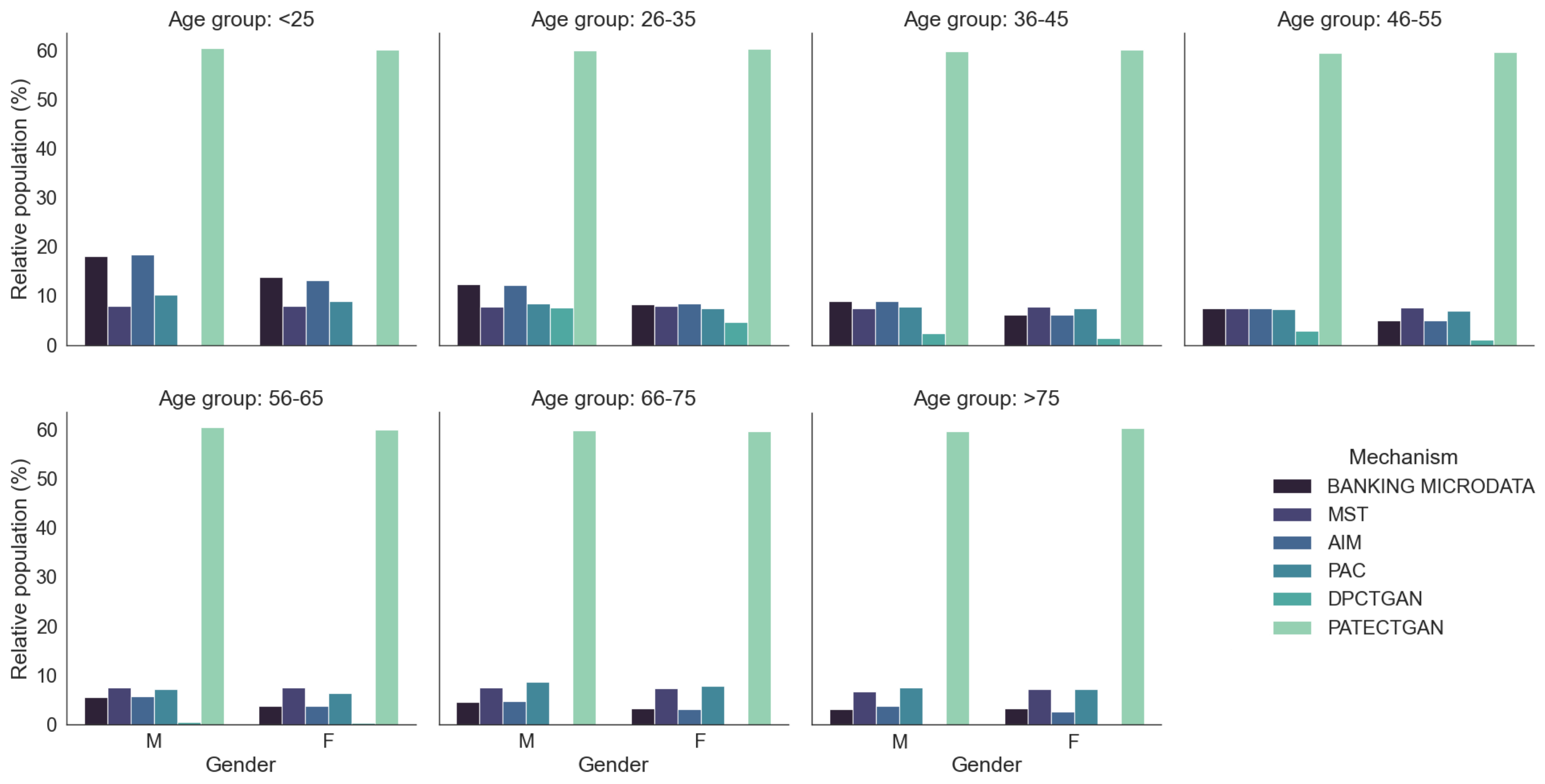}
    \caption{Delinquency rate of credit cards across age and gender groups, based on 2023 (original) banking and synthetic microdata using CBP's PS. Missing values correspond to levels without generated data.}
    \label{fig:figure_4}
\end{figure}

In this application, we transform credit card microdata to improve statistical disclosure with demographic and loss data. Only cards active in both 2020 and 2021 are considered to focus the transition matrices on delinquency states, which results in a 80\% and 87\% coverage with respect to the total number and debt of credit cards, respectively. One benefit of using SDG is that loss data can be extended to age and gender. Figure~\ref{fig:figure_4} shows that men and younger age groups have higher delinquency rates, which is an insight not currently obtainable through other statistical disclosure methods at the CBP. Similarly, loss data is incorporated into the CBP's statistical disclosure as transition matrices for delinquency and debt states. 

The delinquency transition matrix shows that at a delinquency of 90--120 delinquency days, default became the most probable state in a year, and that a credit card's debt level tends to decrease or maintain its previous level, which could be explained by the limit of a card's line of credit. Eventually, synthetic banking microdata released in this area could consider an individual's collection of cards to produce their comprehensive debt level.  

We transform the original banking microdata using the data-driven PS to evaluate whether data-driven binning techniques defined in Table ~\ref{tab:usecase3_features} yield different results. The transition matrices show that the data-driven pre-processing strategy results in more stable delinquency states. This is not surprising considering that delinquency days are binned using $k$-means over a range of 8,000 days, but delinquency could only increase by 365 days per year. 

\begin{table}[!htb]
\caption{Frobenius Norm $\parallel M_{m} \parallel_F$ of the difference between transition matrices of original and synthetic credit card data for 2020 and 2021. `deb' and `del' represent debt and delinquency transition matrices, respectively.}
\begin{footnotesize}
\begin{center}
\begin{tabular}{ l r r c r r }
\toprule
\multicolumn{1}{c}{} & \multicolumn{2}{c}{\textbf{CBP}} & \multicolumn{1}{c}{} & \multicolumn{2}{c}{\textbf{Data-driven}}\\
\cline{2-3} \cline{5-6}
\textbf{Mechanism} & $\parallel M_{m} \parallel_{F, \mathrm{deb}}$ & $\parallel M_{m} \parallel_{F, \mathrm{del}}$ & & $\parallel M_{m} \parallel_{F, \mathrm{deb}}$ & $\parallel M_{m} \parallel_{F, \mathrm{del}}$\\
\midrule
MST&$0.00$&$0.44$& &$0.01$&$0.25$\\
AIM&$0.00$&$0.84$& &$0.01$&$0.20$\\
PAC&$1.11$&$1.56$& &$1.03$&$1.19$\\
DPCTGAN&$0.72$&$2.27$& &$0.73$&$3.15$\\
PATECTGAN&$1.29$&$1.59$& &$1.37$&$2.21$\\
\bottomrule
\end{tabular}
\label{tab:Frobonius errors}
\end{center}
\end{footnotesize}
\end{table}

Similar to our yield curve application, one challenge of the data-driven PS is discretisation, which reduces the information product's utility in understanding loss dynamics due to large bin widths. In terms of total debt, almost all bins cover a limited range of 0--7 million PYG because a large proportion of credit cards contains a small amount of debt. The synthetic data error is demonstrated for both strategies through the relative delinquency in Figure \ref{fig:figure_4} and the Frobenius norm of the difference matrix in Equation \ref{eq:fro}, between the mechanisms and the original banking microdata, in Table ~\ref{tab:Frobonius errors}. Again, MBI mechanisms outperform GANs in both aspects and pre-processing strategies, which supports the utility for domain applications based on frequency tables. 

\section{Discussion}

We demonstrate synthetic banking microdata to be a viable alternative for financial regulators wishing to augment their banking microdata statistical disclosure, but attention should be given to how utility metrics will be used. First, domain applications reveal that the utility of synthetic banking microdata is very sensitive to the domain PS employed. Information products that use numerical features to compute metrics such as weighted averages are especially vulnerable, and a trade-off exists between model performance and feature relevancy. 

For example, the term yield curve application highlights how inadequate binning of features, such as defining bins with missing data, results in poor synthetic data. Using data-driven discretisation like as equal-frequency or $k$-means binning improves the fidelity of synthetic data but results in poor relevancy. On the other hand, information products that rely on frequency tables, such as the usage financial inclusion index, are not exposed to post-processing information loss. 

Secondly, this work emphasises the importance of evaluating SDG utility at each domain application's intended granularity level. Each application demonstrates how utility error metrics propagate unevenly between feature classes, limiting the accuracy that aggregated statistics have on specific relations. Understanding the limitations of domain applications and mechanisms is critical for financial regulators because poor preservation of pair relation between features, such as FI and demographic variables, may lead to misfocussed policy and regulation decisions. 

Thirdly, domain pre-processing acts as a utility-fidelity hyperparameter that should be tuned between domain and data-driven strategies. The key strength of a domain-based PS is the use of data definitions that are widely adopted by a country's financial system, which results in more utility in terms of the relevancy and comparability. For example, the delinquency categories used in transition matrices are well-understood by banks, lending institutions, and financial users. 

Data-driven pre-processing strategies, however, show improved modelling in terms of range coverage and utilise the underlying feature distribution more effectively. For example, equal-frequency binning handles the missing data of the CBP's current grouping for term deposit maturity and significantly improves the synthetic yield curves. 

Additionally, the domain applications demonstrate how synthetic banking microdata is effective in expanding the granularity of existing statistical disclosure, especially through demographic variables that are not currently accessed. However, the motivation to match public datasets should be approached carefully. Demographic variables are commonly used in other public datasets, such as Paraguay's national statistics agency, and it might be tempting to link datasets. 

One could, for example, replicate age groups of synthetic data to match those used in labour statistics and combine datasets to explore banking and economic variables. While microdata exclusively reports on banked individuals, labour statistics include the unbanked. Another potential issue with data provenance in the FI index is that survey data used in the GFD defines the active population as age 15 and above, while banking microdata only starts at 18. 

Regulators should continuously evaluate privacy and disclosure-associated risks with regard to each domain application. In this exploratory work, the same strict privacy parameters are used for all applications, but in practice, there are different privacy requirements. For example, term deposit features such as capital are very sensitive, and the CBP uses savings microdata very conservatively, while the total number of credit cards by age group may not represent the same potential risk, as it does not disclose personally identifiable information. 

Using the same parameters in both applications equates the potential harm of disclosing an individual's ownership of a credit card and the amount invested in a term deposit. There are also practical considerations that regulators face when implementing SDG, especially in developing countries. Financial regulators with more experience in data science and machine learning applications with banking microdata will have more comprehensive and effective data pipelines. 

However, central banks that mostly use banking microdata for aggregated reporting, such as the Central Bank of Paraguay, could find issues with data quality as data validation might be missing at more granular levels. This is amplified in historical banking data, when controls were not in place and suitable data imputation techniques are necessary. There is also a difference between financial regulators in terms of available computational power. Many regulators, such as the CBP, may not have private clouds or GPU clusters that are readily available for SDG and cannot use public clouds because of confidentiality requirements.  

One limitation of this paper's scope as an application-oriented piece of research is that, as opposed to broad benchmarking works such as \citet{tao2021benchmarking}, not all available SDG mechanisms can be reasonably tested. The data used to generate synthetic banking microdata is also encoded to categorical features, which could adversely affect mechanisms that work with numeric features. Binning numerical features may provide additional privacy preservation, but it is expected that the final versions of synthetic microdata releases should maintain the original properties.  

Clear directions of future work are mentioned throughout the work, but there are two main issues we believe should be prioritised when considering the relevance of statistical disclosure for financial regulators and the impact of new, synthetic information products on policy. First, improved pre-processing strategies that optimise the utility of each application should be explored. This work shows that utility metrics can be used to tune a PS as a hyperparameter, meaning the number of bins and discretisation techniques. Pre-processing strategies can be enhanced with additional partitioning algorithms using more of the feature space, such as clustering methods. 

Secondly, post-processing is a major determinant of the utility of synthetic banking microdata and should be addressed within an SDG strategy. The domain applications used in this work demonstrate that information products that require numeric features to calculate relevant metrics and use simple post-transformation functions, unsurprisingly, yield limited results. More robust techniques stand a chance to deliver improved utility and fidelity. For example, the 1-way marginal of each pre-processed feature level could be maintained and later used for decoding.

\section{Conclusion}

This work presents the main findings of an initial exploration of synthetic data generation for financial regulators using the Central Bank of Paraguay's banking microdata. In doing so we provide the, to our knowledge, first open-data release of synthetic data generated directly from a full central bank dataset. Three representative domain applications are implemented to evaluate such data through information products connected to a financial regulator's priorities. 

We include demographic variables for every application to leverage the data's granularity, and each application's utility is evaluated through error metrics, providing a comprehensive view of the error propagation. We also assess two separate pre-processing strategies, one following the CBP's current data definitions and one data-driven. Our results show that SDG is a promising technology to complement statistical disclosure. We find that utility metrics for domain applications should be evaluated at their intended granularity, as errors may be distributed unevenly between granularity levels. Such considerations need to be clearly expressed in a financial regulator's SDG strategy, which should explain the different trade-offs between utility, fidelity, and privacy. 

Our data extraction, validation, and processing indicate that regulators with less experience and resources are likely to face additional challenges. Experiments also show that domain applications that do not incur post-processing information loss and are based on frequency tables with well-represented classes are better suited for SDG. The utility of the synthetic data supports the consensus that marginal-based inference mechanisms outperform GANs for tabular data. 

In cases with adequate utility, our analysis leads to concrete information products that provide more granular data than current statistical disclosure. For example, the loss data for credit cards is information not currently featured and valuable to financial institutions, researchers, and practitioners. We also identify priorities for further research to implement an SDG strategy at the CBP. The synthetic datasets are publicly released to encourage further dialogue between financial regulators, privacy practitioners, and experts in statistical disclosure. 

Synthetic banking microdata generation can be used to nurture and maintain the public trust that central banks needs to work effectively. In turn, this helps financial regulators to mitigate reputational risks associated with information leakage and build trust through explainable, transparent data governance. SDG has the potential to transform the way central banks interact with financial markets and contribute to more accountable and resilient public institutions. 

\bibliographystyle{apalike}
\bibliography{cas-refs.bib}

\end{document}